\documentclass[preprint,superscriptaddress,showpacs,aps]{revtex4}

\usepackage{graphicx}
\usepackage{subfigure}
\usepackage{amsmath}
\usepackage{amssymb}

\begin{document}

\title{Artificial electromagnetic field with cold atoms in two-dimensional optical lattice}

\author{Xiaoyong Guo}
\email{gxyauthor@tust.edu.cn}
\affiliation{School of Science, Tianjin University of Science and Technology, Tianjin 300457, China}

\author{Yu Chen}
%\email{spaceexplorer@163.com}
\affiliation{Center for Theoretical Physics, Department of Physics, Capital Normal University, Beijing, 100048, China}

\author{Jian Liu}
%\email{liujian@upc.edu.cn}
\affiliation{College of Science, China University of Petroleum (East China), Qingdao 266580, China}
%%%%%%%%%%%%%%%%%%%%%%%%%%%%%%%%%%%%%%%%%%%%%%%%%%%%%%%%%%%%%%%%%%%%%%%%%%%%%%%%%%%%%%%%%%%%%%%%%%%%%%%%
\begin{abstract}

We propose that an artificial electromagnetic field can be engineered in the context of cold fermionic atoms that are coupled to a cavity mode via two-photon processes in a two-dimensional optical lattice. There is a standing-wave pump laser inducing the electric effect in one spatial direction, and a second running-wave laser beam generates the magnetic flux perpendicular to the lattice plane. In the static scenario, the bulk spectrum resembles the fractal structure of the Hofstadter butterfly and the edge mode spectrum indicates the occurrence of the quantum Hall phase. The Keldysh formulism is utilized to capture the time evolution. The back action between atoms and cavity field gives a picture of the time-dependent non-equilibrium dynamics. We find that the spontaneous emergence of the artificial electromagnetic field stimulates the Hall current, and the superradiant cavity field emerges without pump threshold. Moreover, the magnetic flux not only modify the bulk topology, but also drives a series of dynamical phase transitions.

\end{abstract}


\maketitle

%%%%%%%%%%%%%%%%%%%%%%%%%%%%%%%%%%%%%%%%%%%%%%%%%%%%%%%%%%%%%%%%%%%%%%%%%%%%%%%%%%%%%%%%%%%%%%%%%%%%
\section{Introduction}

More than thirty years ago, a vision of using one quantum mechanical system to emulate the physics of a computationally or analytically intractable one had been proposed \cite{Feynman}. Nowadays, this idea has inspired tremendous research activities on the subject of quantum simulators. Owing to the well controllable microscopic physics \cite{Gorlitz,Chin}, cold atomic gases are ideal platform for investigating challenges and open questions that may originally be encountered in the context of traditional condensed matter physics or high-energy physics. Excellent examples include the Bardeen-Cooper-Schrieffer (BCS) to the Bose-Einstein condensation (BEC) crossover in degenerate Fermi gases \cite{nsr,Greiner1,Regal,Giorgini}, the bosonic superfluid to Mott transition in optical lattice \cite{Greiner,Larson}, and the rich phase diagram of the Bose-Hubbard model \cite{Bakhtiari,Klinder,Landig,Chen2}.
In hindsight, the underlying physics of these examples is governed by the short-range interactions. A feasible way to engineer long-range correlations is to couple the atoms to an optical cavity, in which all atoms can be dipole interacted with a common cavity mode making the effective interaction infinitely long-ranged \cite{Asboth,Asboth1}. This celebrating advance makes quantum degenerate gases highly versatile for creating long-range physics, such as the phase transition from an atomic BEC to a supersolid phase \cite{Mottl,Gopalakrishnan} and the Dicke quantum phase transition from normal to superradiance \cite{Baumann,Nagy,Chen,Chen1}.

Besides the long-range interactions, introducing an optical cavity into the field of quantum emulators has another advantage: generating a dynamical gauge field. It is well known that atoms are charge neutral, and therefore, their motion is not affected by the external electromagnetic field. However, the matter-light interaction offers a possibility to create various gauge potentials that modify the properties of the atoms \cite{Goldman,Spielman}. Using this approach, atoms may emulate the motion of charged particles subjected to the Abelian or non-Abelian gauge potentials. In fact, most of proposed setups concern the realization of background and classical gauge potentials. As such, they are not influenced by the matter field and need not obey Maxwell's equations \cite{Dalibard}. Consequently, a theory of artificial gauge field with dynamical nature is greatly desirable.
Recent works \cite{ZhengW,Kollath} demonstrate that the dynamical Abelian gauge fields can be designed in an optical lattice by the Raman process between a pump laser and a single mode cavity. The atoms and the cavity field possess their own quantum degrees of freedom that are dynamically related via the matter-light interaction.
In Ref. \cite{ZhengW}, a synthetic electric field is fabricated in a one-dimensional (1D) optical lattice, where the native hopping that is forbidden by the energy offset is restored via a two-photon process. When the cavity occupation subjects to the leakage, the atoms are self-organized into a state with unidirectional movement. The resulting chiral atomic flow mimics the motion of charged particles under an uniform electric field.
On the other hand, in Ref. \cite{Kollath}, a similar coupling mechanism realizing a dynamical magnetic field is proposed. In the geometry of an array of two-dimensional (2D) decoupled ladders, a Raman process between a running-wave pump beam and an optical cavity that bridges two legs of a single ladder is induced. In the mean time, the native tunneling along the legs is allowed. Owing to the running-wave nature of the pump beams, a site-dependent phase is imprinted. After the completion of a trajectory around a single plaquette, the atomic wavefunction will acquire a phase factor. An artificial magnetic field perpendicular to the ladder plane is dynamically emerged, since the cavity field is a dynamical degrees of freedom. In the steady state, either a chiral insulator or a chiral liquid carrying chiral edge currents are formed on the legs.

Inspired by the progresses mentioned above, it is curious to seek how to emulate a fully dynamical gauge field which includes electric and magnetic effects simultaneously. In this proposal, the atoms and the gauge field could have their own quantum degrees of freedom, and the back action between them will sharp the dynamical behavior. On the other aspect, the system may be characterized by non-equilibrium dynamics due to the cavity loss. Hence it is quite natural to surmise that when we successfully design such a setting, it will shed some light on realizing quantum electrodynamics with cold atoms and even on the non-equilibrium dynamics of the coupled matter-light system.
In this paper, we address these open questions by proposing a 2D lattice model where matter and light are coupled via the two-photon processes between pump lasers and a single cavity mode. In addition to a standing-wave pump beam, we introduce a running-wave laser beam. We argue that our model can realize the dynamical electric and magnetic effects in a configuration mimicking the Hall system of 2D electron lattice. The Harper equation and the Keldysh formalism are employed. The realization of the quantum Hall phase, emergence of the superradiance, and temporal behavior of atomic currents are discussed.

The paper is organized as follows. In the subsequent section, the model Hamiltonian and its physical realization are discussed. By the associated Harper equation, the bulk and edge mode spectra are calculated. In Sec. \ref{dynamics}, the equation of motion and its numerical solutions are provided. The currents describing the particle flow and the emerging of self-organized electromagnetism are revealed. We summarize the new findings which do not present in the previous works in Sec. \ref{con}. Throughout this paper, we use natural units $\hbar=c=1$.
%%%%%%%%%%%%%%%%%%%%%%%%%%%%%%%%%%%%%%%%%%%%%%%%%%%%%%%%%%%%%%%%%%%%%%%%%%%%%%%%%%%%%%%%%%%%%%%%%%%%
\section{model}
\label{model}

%%%%%%%%%%%%%%%%%%%%%%%%%%%%%%%%%%%%%%%%%%%%%%
\begin{figure}[tb]
\centering
\includegraphics[width=12cm]{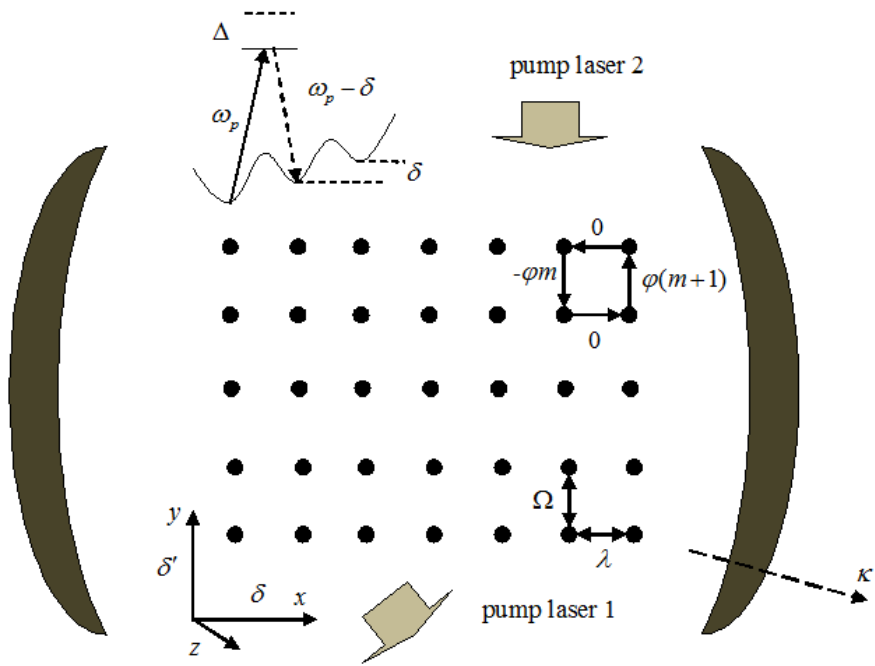}
\caption{Setup for dynamical artificial electromagnetic field. Neighbouring sites are shifted in energy by the offsets $\delta$ and $\delta'$ so as to inhibit the natural hopping. Two cavity-assisted Raman processes then allow to reestablish the hopping in a controllable manner. The first one with the atom-laser coupling $\lambda$ generates an artificial electrical field along $x$-direction, while the other with coupling $\Omega$ effects an artificial magnetic field along $z$-direction with $\phi$ being the flux through a single plaquette. The cavity decay is denoted by $\kappa$.}
\label{fig1}
\end{figure}
%%%%%%%%%%%%%%%%%%%%%%%%%%%%%%%%%%%%%%%%%%%%%%%%%%%%%%%%%%%
The schematic of our coupling strategy is shown in Fig. \ref{fig1}. The key ingredients are elaborated in this section. We suppose an optical cavity with a single standing-wave cavity mode with frequency $\omega_{c}$. The cavity field is oriented in the $x$-direction, and the intra-cavity photons may be leaked out via the cavity decay rate $2\kappa$.
The atomic center-of-mass motion is restricted to the $x-y$ plane by a square optical lattice that freezes the $z$ degree of freedom. The lattice spacing $d=1$ is chosen as the unit of length throughout this text. The lattice size is $L\times L$, where $L$ is the number of sites in each orientation. By accelerating the optical lattice or applying a real gradient magnetic field, the energy offsets $\delta$ and $\delta'$ can be imposed along the positive $x$ and $y$-axes respectively. So that, the lattice is energetically tilted along $\mathbf{e}_{x}+\mathbf{e}_{y}$, where $\mathbf{e}_{i}$ denote the unit vectors with $i=x,y,z$. The lattice's native hopping can then be suppressed by assuming the large value of $\delta$ and $\delta'$.
The first standing-wave pump laser is linearly polarized and it incident along the $z$-direction with frequency $\omega_{p}$. With the help of this beam, an atom can hop to the right neighboring site by absorbing a laser photon and then emitting it to the cavity mode at frequency $\omega_{p}-\delta$, which is detuned from the cavity resonant frequency by $\Delta=\omega_{c}-\omega_{p}+\delta$. Thus, the natural tunneling, which is stymied by the energy offset $\delta$, is restored via this cavity-assisted Raman process. However hopping to the opposite is impossible in this process, because we may assume that $\delta$ is sufficiently large, that emission at $\omega_{p}+\delta$ is negligible. As a result, such a process can push the atoms to moving on the positive $x$-axis acting like a transverse voltage for the charge particles.
In addition, a second running-wave pump laser is incident with frequency $\omega_{p}'=\omega_{c}+\delta'-\Delta$ and wave-vector $\mathbf{k}_{p}=k_{p}\mathbf{e}_{y}+k_{p,z}\mathbf{e}_{z}$. It is also linearly polarized. This laser beam has two functions. The one is that it generates a similar Raman process as the first pump laser realizing the cavity-assisted Raman process along $\mathbf{e}_{y}$. The other is that it imprints a spatially dependent phase factor $e^{\mathrm{i}\phi m}$ onto the this process. Here, the phase factor $\phi=k_{p}d$ depends on the incident angle of this laser beam. The effect of the artificial magnetic field can be seen if an atom tunnels once around a plaquette. Particularly, hopping from the site $(m,n)$ to $(m,n-1)$, it will collect a phase $-\phi m$. However, jumping in the $\mathbf{e}_{x}$ does not acquire any additional phase. Finally, when it jumps back from site $(m+1,n-1)$ to $(m+1,n)$, a phase $\phi(m+1)$ is picked up. Therefore, a net phase factor $\phi$ is acquired when the atom returns to the initial site $(m,n)$, corresponding to a magnetic field $\mathbf{B}$ in the Landau gauge. One can reach the strong magnetic field regime, where the phase $\phi$ can take any value between $0\rightarrow2\pi$.

With all these ingredients in hand, the setup can generate an artificial electric field in the $x$- direction, and in the mean time an artificial magnetic field in the $z$-direction. In this sense, our system mimics the charged particles in the Hall system. Under the tight-binding approximation, the effective Hamiltonian in a rotating frame at the frequency $\omega_{p}-\delta$ can be written as
\begin{equation}\label{ham1}
\begin{aligned}
H=&\Delta a^{\dag}a-\lambda\sum_{mn}\left(a^{\dag}c^{\dag}_{m+1,n}c_{m,n}
+ac^{\dag}_{m,n}c_{m+1,n}\right)\\
&-\Omega\sum_{mn}\left(e^{\mathrm{i}2\pi m\Phi}a^{\dag}c^{\dag}_{m,n+1}c_{m,n}
+e^{-\mathrm{i}2\pi m\Phi}ac^{\dag}_{m,n}c_{m,n+1}\right),
\end{aligned}
\end{equation}
where, $a$ is the annihilation operator for the cavity photon, $c_{m,n}$ is annihilation operator for fermionic atoms at site $(x,y)=(md,nd)$ with $(m,n)\in\mathbb{Z}$, $\Phi=\phi/2\pi=\Phi_{0}^{-1}\int \mathbf{B}\cdot d\mathbf{S}$ is the number of magnetic flux quanta $\Phi_{0}=2\pi$ penetrating the plaquette, and the intensities of the atom-laser couplings are denoted by $\lambda$ and $\Omega$. Since in the Hamiltonian (\ref{ham1}) the cavity field is not a static external potential but a dynamical quantum degrees of freedom, the artificial electric and magnetic fields may reproduce a complete field-theory picture, in which they have their own dynamics.

When we replace the dynamical cavity field $a$ by an additional laser field $\alpha=|\alpha|e^{-\mathrm{i}\theta}$, a model Hamiltonian with static gauge potential is obtained,
\begin{equation}\label{hamstatic}
\begin{aligned}
H=&\Delta |\alpha|^{2}-\lambda|\alpha|\sum_{mn}\left(e^{\mathrm{i}\theta}c^{\dag}_{m+1,n}c_{m,n}
+e^{-\mathrm{i}\theta}c^{\dag}_{m,n}c_{m+1,n}\right)\\
&-\Omega|\alpha|\sum_{mn}\left(e^{\mathrm{i}(2\pi m\Phi+\theta)}c^{\dag}_{m,n+1}c_{m,n}
+e^{-\mathrm{i}(2\pi m\Phi+\theta)}c^{\dag}_{m,n}c_{m,n+1}\right).
\end{aligned}
\end{equation}
Notice that for a laser field of real amplitude (i.e., $\theta=0$), the above Hamiltonian reduces to the familiar Hofstadter model \cite{Hofstadter} with modulated hopping matrix elements $\lambda|\alpha|$ and $\Omega|\alpha|$. Therefore, Eq. (\ref{hamstatic}) can be regarded as a generalization of the Hofstadter model. In the rest of this section, we shall address the quantum Hall phase with topological order of this model by calculating the bulk spectrum and the edge spectrum. The single-particle eigenstate of the Hamiltonian (\ref{hamstatic}) can be searched in the form $|\Psi\rangle=\Sigma_{mn}e^{\mathrm{i}\nu n}\psi_{m}c^{\dag}_{m,n}|0\rangle$, where $|0\rangle$ is the vacuum state. The corresponding Harper equation for the coefficients $\psi_{m}$ reads
\begin{equation}\label{Harper}
\frac{E}{|\alpha|}\psi_{m}=-\lambda\left(e^{\mathrm{i}\theta}\psi_{m-1}
+e^{-\mathrm{i}\theta}\psi_{m+1}\right)-2\Omega\cos(2\pi m\Phi-\nu+\theta)\psi_{m}.
\end{equation}
In the discussion below, all energies are in units of $|\alpha|$. Besides $|\alpha|$, there are four adjustable parameters: $\lambda$, $\Omega$, $\Phi$, and $\theta$. It is impossible to show all the results from the whole four-dimensional parameter space. Therefore, we selectively report on certain sets of parameters with representative features.
%%%%%%%%%%%%%%%%%%%%%%%%%%%%%%%%%%%%%%%%%%%%%%
\begin{figure}[tb]
\centering
\includegraphics[width=15cm]{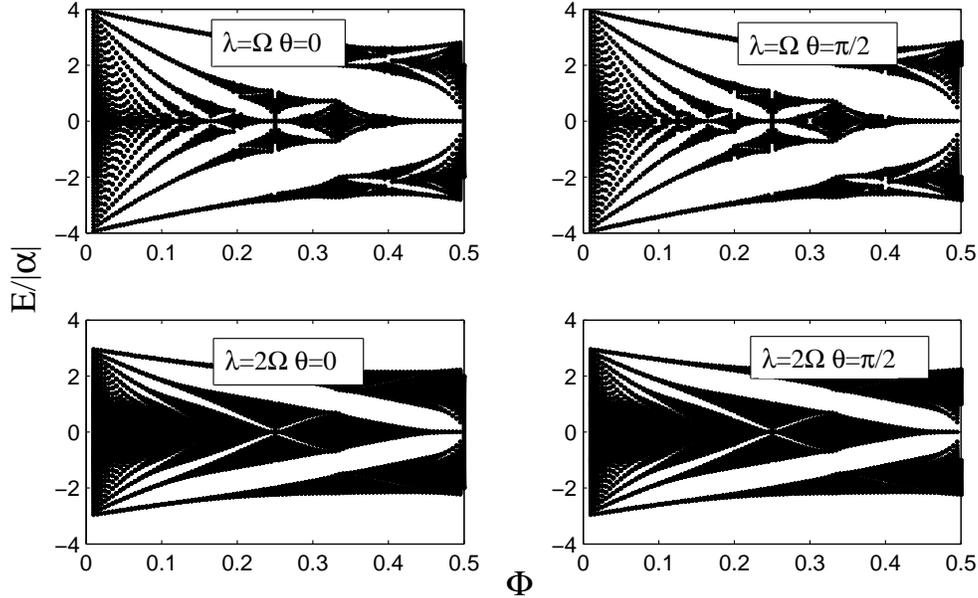}
\caption{Bulk energy spectrum as a function of magnetic flux for various parameters.}
\label{fig2}
\end{figure}
%%%%%%%%%%%%%%%%%%%%%%%%%%%%%%%%%%%%%%%%%%%%%%%%%%%%%%%%%%%
%%%%%%%%%%%%%%%%%%%%%%%%%%%%%%%%%%%%%%%%%%%%%%
\begin{figure}[tb]
\centering
\includegraphics[width=15cm]{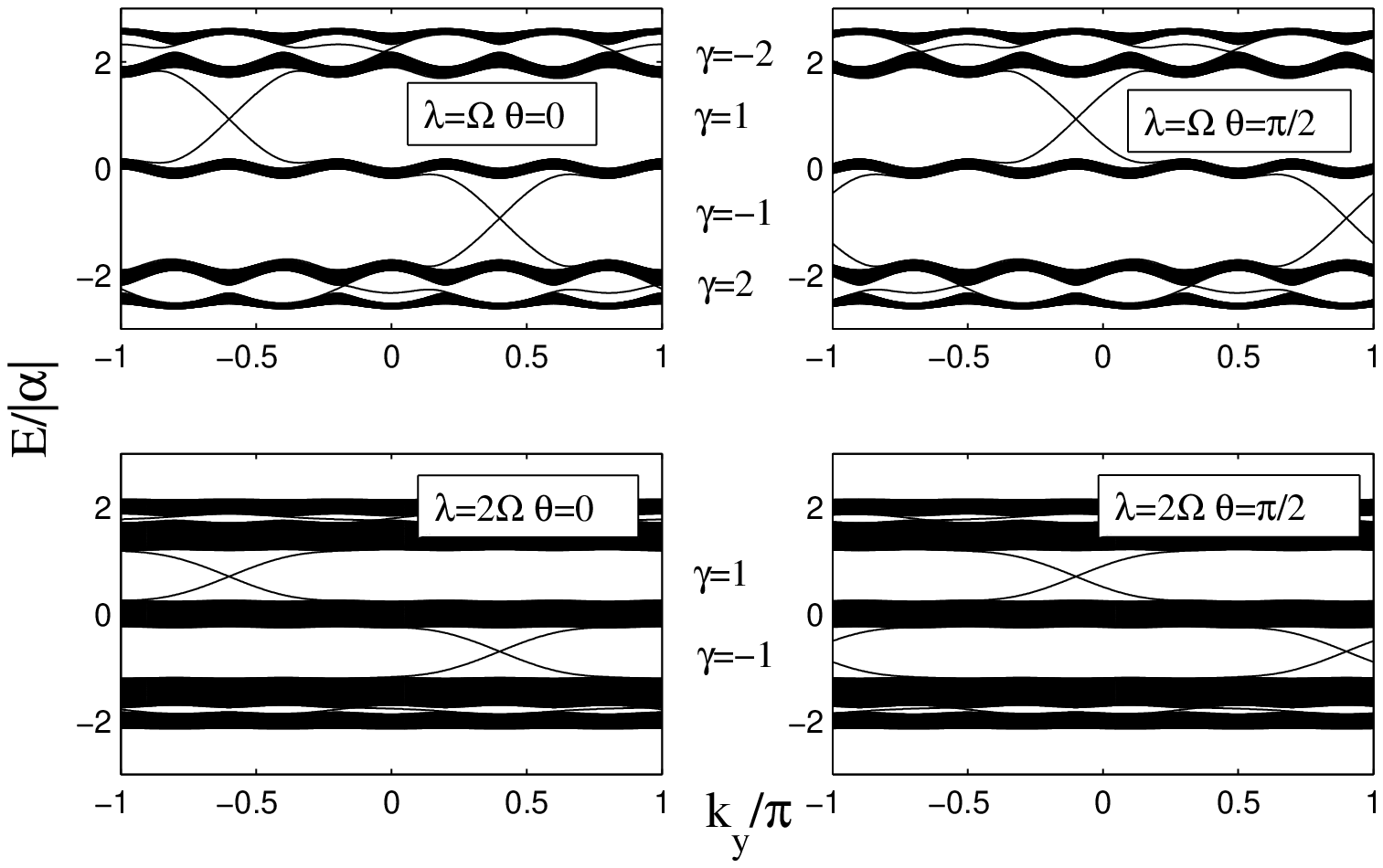}
\caption{Edge mode spectrum as a function of quasi-momentum for $\Phi=2/5$. Here, $\gamma$ is the topological invariant for each bulk gap. Other parameters in each panel are the same as in Fig. \ref{fig2}.}
\label{fig3}
\end{figure}
%%%%%%%%%%%%%%%%%%%%%%%%%%%%%%%%%%%%%%%%%%%%%%%%%%%%%%%%%%%

To obtain the bulk energy spectrum, we solve Eq. (\ref{Harper}) on a toroidal geometry by considering the periodic boundary condition in both spatial directions. The results are plotted as a function of the magnetic flux in Fig. \ref{fig2}. For the simplest case of equal matter-light coupling $\lambda=\Omega$ with a real laser field $\theta=0$, the spectrum reproduces the well known Hofstadter butterfly as provided in \cite{Hofstadter}. We note that the only way that the laser phase $\theta$ effects in Eq. (\ref{Harper}) is additive to the quantity $\nu$. Thus the fractal structure of the energy spectrum can not be changed by this laser field. In the lower panels of Fig. \ref{fig2}, we plot the cases of $\lambda=2\Omega$. It is shown that the unequal matter-light couplings alter the fractal structure. Particularly, certain bulk gaps that are open in the upper panels within the interval $0.3<\Phi<0.5$ are closed.

The edge mode analysis can be performed by solving Eq. (\ref{Harper}) on a cylinder geometry, where the open boundary condition in $x$-direction and periodic boundary condition in $y$-direction are applied. In this scenario, the quantity $\nu$ should be identified as the quasi-momentum $k_{y}$ along the $\mathbf{e}_{y}$ axis. We present the results in Fig. \ref{fig3} where the magnetic flux is $\Phi=2/5$, i.e. in the interval $0.3<\Phi<0.5$, and other parameters in each panel are identical to Fig. \ref{fig2}. In the original Hofstadter model, when the flux is rational $\Phi=p/q$, where $p$ and $q$ are integers, the spectrum will display $q$ bulk bands with $q-1$ gaps. The similar situation is true for the model (\ref{hamstatic}) with equal matter-light couplings. In the upper panels, the spectrum consists of four gaps within five bulk bands. However, in the lower panels there are two gaps within three bulk bands. As we already mentioned, the unequal couplings manipulate fractal structure of the Hofstadter butterfly by closing certain bulk gaps.
Indeed, we typically find a few edge states within the bulk gaps, some of which cross the gap from one bulk band to the other. Importantly, each edge state contributes $e^{2}/2\pi$ to the Hall conductivity of the system. In order to evaluate the Hall conductivity of a bulk gap, we count the edge states whose dispersions intersect the Fermi energy, taking into account their location and direction of the group velocity $\partial E(k_{y})/\partial k_{y}$. Therefore to the Hall conductivity reads
\begin{equation}\label{Hallconductivity}
\sigma_{H}=e^{2}/2\pi\gamma,
\end{equation}
where $\gamma=N_{R}-N_{L}$ with $N_{L}$ and $N_{R}$ being the number of left- and right-moving states, respectively. Notice that the quantity $\gamma$ is just the first Chern number for each bulk gap. The calculated $\gamma$ is also marked in Fig. \ref{fig3}. The laser phase influences the profile of the edge modes, but the topology is governed by the matter-light couplings. It is shown that for the case $\lambda=\Omega$, the four gaps are characterized by four different Chern numbers. When the two couplings are unequal, two gaps with $\gamma=\pm2$ are closed and the edge modes in these gaps are merged into bulk bands. The other two gaps with $\gamma=\pm1$ remains open and the corresponding edge modes are intact. As a result, a topological phase transition is induced by the varying matter-light couplings.
%%%%%%%%%%%%%%%%%%%%%%%%%%%%%%%%%%%%%%%%%%%%%%%%%%%%%%%%%%%%%%%%%%%%%%%%%%%%%%%%%%%%%%%%%%%%%%%%%%%%
\section{Non-equilibrium dynamics}
\label{dynamics}

The non-equilibrium dynamics of the system is described by the master equation $\partial_{t}\rho=-\mathrm{i}\left[H,\rho\right]+\mathcal{L}\rho$, where $\rho$ is the reduced density matrix, and the cavity loss is encoded in the Lindblad dissipator $\mathcal{L}\rho=\kappa\left(2a\rho a^{\dag}-a^{\dag}a\rho-\rho a^{\dag}a\right)$ \cite{Orszag}. For the Hamiltonian (\ref{ham1}), the equation of motion of the average $\alpha=\langle a\rangle$ reads
\begin{equation}\label{eomp}
\partial_{t}\alpha=-\mathrm{i}(\Delta-\mathrm{i}\kappa)\alpha+\mathrm{i}\sum_{ij}
\left(\lambda\langle c^{\dag}_{i+1,j}c_{i,j}\rangle+\Omega e^{\mathrm{i}2\pi i\Phi}\langle c^{\dag}_{i,j+1}c_{i,j}\rangle\right).
\end{equation}
This equation describes the temporal evolution of the coherent cavity field. On the other hand, the equation of motion of the atomic density matrix, $\rho_{iji'j'}(t)=\langle c^{\dag}_{i,j}(t)c_{i',j'}(t)\rangle$, can be obtained by the virtue of the Keldysh formulism \cite{Rammer,Altland,Sieberer}. After some algebra, we find that
\begin{equation}\label{eomfull}
\begin{aligned}
\partial_{t}\rho_{iji'j'}(t)=&\mathrm{i}\lambda(\alpha^{\ast}\rho_{iji'-1j'}+\alpha\rho_{iji'+1j'}
-\alpha^{\ast}\rho_{i+1ji'j'}-\alpha\rho_{i-1ji'j'})\\
&+\mathrm{i}\Omega(\alpha^{\ast}e^{\mathrm{i}2\pi i'\Phi}\rho_{iji'j'-1}
+\alpha e^{-\mathrm{i}2\pi i'\Phi}\rho_{iji'j'+1}
-\alpha^{\ast}e^{\mathrm{i}2\pi i\Phi}\rho_{ij+1i'j'}
-\alpha e^{-\mathrm{i}2\pi i\Phi}\rho_{ij-1i'j'})\\
&-\frac{\kappa}{\Delta^{2}+\kappa^{2}}\sum_{\alpha\beta ml}\left[
v_{1}(i'j'\alpha\beta)v_{2}(\alpha\beta ml)\rho_{ijml}
+v_{1}(ml\alpha\beta)v_{2}(\alpha\beta ij)\rho_{mli'j'}\right]\\
&+\frac{2\kappa}{\Delta^{2}+\kappa^{2}}\sum_{\alpha\beta\alpha'\beta'}
v_{2}(i'j'\alpha\beta)v_{1}(\alpha'\beta' ij)
\rho_{\alpha'\beta'\alpha\beta}\\
&+\frac{\kappa}{\Delta^{2}+\kappa^{2}}\sum_{\alpha\beta\alpha'\beta'}\sum_{ml}\left[
(\bar{v}_{1}-\bar{v}_{2})\rho_{\alpha'\beta'\alpha\beta}\rho_{ijml}
+(\bar{v}'_{1}-\bar{v}'_{2})\rho_{mli'j'}\rho_{\alpha'\beta'\alpha\beta}\right].
\end{aligned}
\end{equation}
Here, the abbreviations are given as $\bar{v}_{1}=v_{1}(i'j'\alpha\beta)v_{2}(\alpha'\beta' ml)$, $\bar{v}_{2}=v_{2}(i'j'\alpha\beta)v_{1}(\alpha'\beta' ml)$, $\bar{v}'_{1}=v_{1}(ml\alpha\beta)v_{2}(\alpha'\beta' ij)$, and $\bar{v}'_{2}=v_{2}(ml\alpha\beta)v_{1}(\alpha'\beta' ij)$, with $v_{1}(iji'j')=\lambda\delta_{ii'+1}\delta_{jj'}+\Omega e^{\mathrm{i}2\pi i\Phi}\delta_{ii'}\delta_{jj'+1}$ and $v_{2}(iji'j')=\lambda\delta_{i+1i'}\delta_{jj'}+\Omega e^{-\mathrm{i}2\pi i\Phi}\delta_{ii'}\delta_{j+1j'}$ being the vertex functions. In the Appendix, we provide the details on the derivation of Eq. (\ref{eomfull}).
To describe the particle flow under the artificial gauge fields, the atomic currents should be defined base on the diagonal elements of the density matrix, $\rho_{ij}=\rho_{ijij}$, together with the continuity equation
$\partial_{t}\rho_{ij}(t)=-(J_{x,ij}-J_{x,i-1j}+J_{y,ij}-J_{y,ij-1})$, where $J_{x}$ and $J_{y}$ are the currents in the transverse and longitudinal orientations. We first defined the coherent current per site as
\begin{equation}\label{cocurrent}
\begin{aligned}
&J^{co}_{x}(ij)=-2\lambda\mathrm{Im}\alpha^{\ast}\rho_{i+1jij}\\
&J^{co}_{y}(ij)=-2\Omega\mathrm{Im}\alpha^{\ast}e^{\mathrm{i}2\pi i\Phi}\rho_{ijij+1}.
\end{aligned}
\end{equation}
It accounts for the atomic hopping by emitting and absorbing coherent cavity photons. Then, following the same logic the classical currents,
\begin{equation}\label{clcurrent}
\begin{aligned}
&J^{cl}_{x}(ij)=-\frac{2\kappa}{\Delta^{2}+\kappa^{2}}\left[
\lambda^{2}\rho_{i+1ji+1j}+\lambda\Omega\cos(\phi i)\rho_{i+1jij+1}\right]\\
&J^{cl}_{y}(ij)=-\frac{2\kappa}{\Delta^{2}+\kappa^{2}}\left[
\Omega^{2}\rho_{ij+1ij+1}+\lambda\Omega\cos(\phi i)\rho_{i+1jij+1}\right],
\end{aligned}
\end{equation}
and the quantum currents,
\begin{equation}\label{qucurrent}
\begin{aligned}
&J^{qu}_{x}(ij)=\frac{2\kappa}{\Delta^{2}+\kappa^{2}}\sum_{ml}\left[
\lambda^{2}\mathrm{Re}\rho_{m+1li+1j}\rho_{ijml}+\lambda\Omega
\mathrm{Re}e^{\mathrm{i}\phi m}\rho_{ml+1i+1j}\rho_{ijml}\right]\\
&J^{qu}_{y}(ij)=\frac{2\kappa}{\Delta^{2}+\kappa^{2}}\sum_{ml}\left[
\lambda\Omega\mathrm{Re}e^{-\mathrm{i}\phi i}\rho_{m+1lij+1}\rho_{ijml}+\Omega^{2}
\mathrm{Re}e^{\mathrm{i}\phi(m-i)}\rho_{ml-1ij+1}\rho_{ijml}\right],
\end{aligned}
\end{equation}
can be found. For the above two equations, the former describes the atomic transport mediated by the photons originated from thermal fluctuations, while the later is for the photons emanating from quantum fluctuations. Since our setup emulates the Hall system, the atomic current along $y$ direction can be interpreted as the analogy of the Hall current, which is defined as $J^{Hall}=\sum_{ij}\left[J^{co}_{y}(ij)+J^{cl}_{y}(ij)+J^{qu}_{y}(ij)\right]$. Whereas, the total transverse current is of $J_{x}=\sum_{ij}\left[J^{co}_{x}(ij)+J^{cl}_{x}(ij)+J^{qu}_{x}(ij)\right]$. The Hall current $J^{Hall}$ is a natural signature of the spontaneous self-organization of the electromagnetic field. In the following, we shall numerically solve Eqs. (\ref{eomp}) and (\ref{eomfull}) discussing the temporal behavior of the cavity field and the particle flow.

In the numerical simulation, the size of the optical lattice is $4\times4$ while other parameters (i.e., couplings $\Omega$ and $\lambda$, and flux $\Phi$) are considered as adjustable. The temporal evolution of the density matrix is determined by the Euler formula $\rho(t+h)=\rho(t)+h\mathbf{f}(\rho(t))$, where the vector $\mathbf{f}$ denotes the left hand side of Eq. (\ref{eomfull}) and $h$ the time step ($h=0.1$ in the simulation). At the beginning $t=0$, the non-diagonal elements of the density matrix $\rho_{i\pm1ji'\pm1j'}(0)$ is set to be non-zero corresponding to a finite initial velocity of the atoms, and in the mean time the cavity is empty $|\alpha|=0$. The initial position of atoms is determined by the non-zero diagonal elements $\rho_{ijij}(0)$. The filling factor $n$ is defined as $n=\frac{N}{L^{2}}$, where $N$ is number of atoms.
%%%%%%%%%%%%%%%%%%%%%%%%%%%%%%%%%%%%%%%%%%%%%%
\begin{figure}[tb]
\centering
\includegraphics[width=12cm]{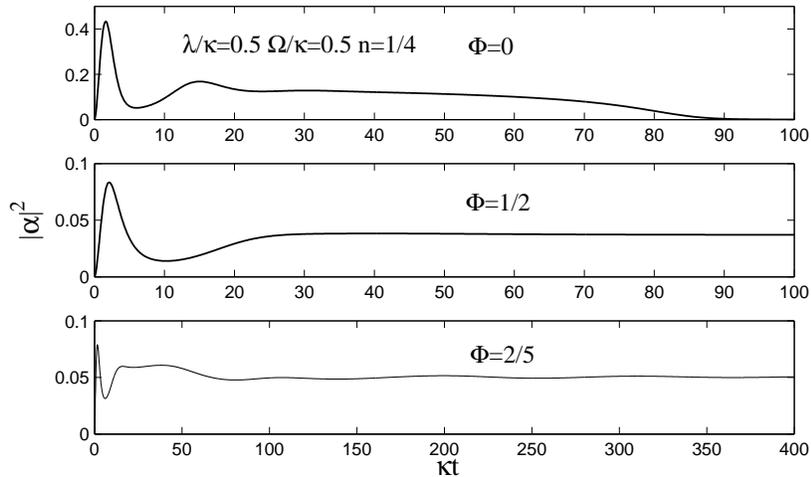}
\caption{Cavity population $|\alpha|^{2}$ including fluctuations plotted as a function of dimensionless time $\kappa t$ for various flux $\phi\theta$. At quarter filling $n=1/4$, the size of the lattice is $L=4$. Other parameters are chosen as $\lambda/\kappa=0.5$, $\Omega/\kappa=0.5$, and $\Delta/\kappa=0.5$.}
\label{fig6}
\end{figure}
%%%%%%%%%%%%%%%%%%%%%%%%%%%%%%%%%%%%%%%%%%%%%%%%%%%%%%%%%%%
%%%%%%%%%%%%%%%%%%%%%%%%%%%%%%%%%%%%%%%%%%%%%%%%%
\begin{figure}[tb]
\subfigure[]%[$b=0.5$]
{
\label{fig7a}
\begin{minipage}[c]{0.5\textwidth}
\centering\includegraphics[width=9.0cm,height=6.3cm]{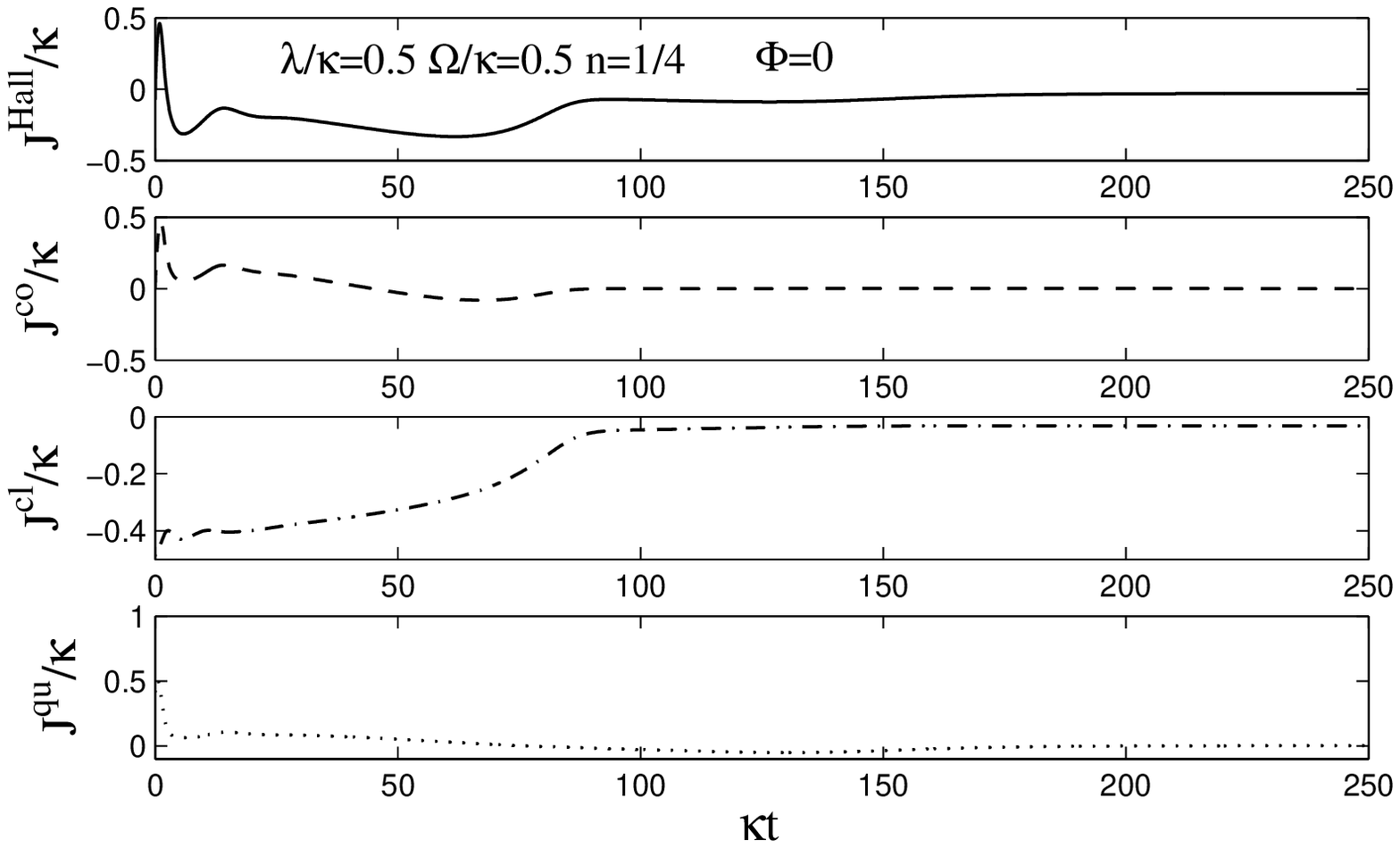}
\end{minipage}}%
\subfigure[]%[$b=3.5$]
{
\label{fig7b}
\begin{minipage}[c]{0.5\textwidth}
\includegraphics[width=9.0cm,height=6.3cm]{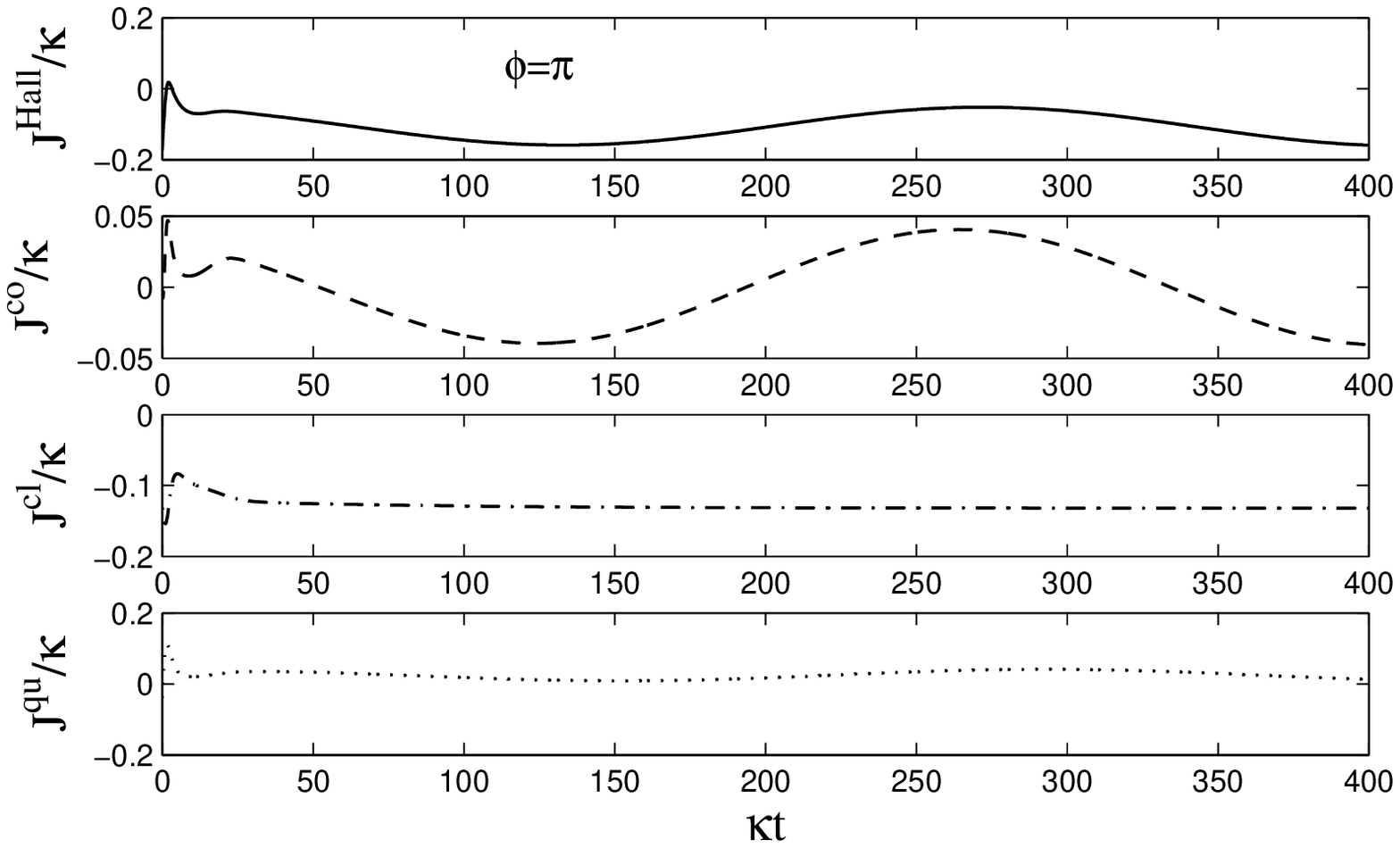}
\end{minipage}}%
\caption{Longitudinal currents $J^{Hall}$, $J^{co}$, $J^{cl}$, and $J^{qu}$ plotted as a function of dimensionless time $\kappa t$. Here, the size of the lattice is $L=4$, and the system is at quarter filling $n=1/4$. Other parameters are chosen as $\lambda/\kappa=0.5$, $\Omega/\kappa=0.5$, and $\Delta/\kappa=0.5$. The magnetic flux is $\Phi=0$ in Fig. \ref{fig7a} and $\Phi=1/2$ in Fig. \ref{fig7b}.}
\label{fig7}
\end{figure}
%%%%%%%%%%%%%%%%%%%%%%%%%%%%%%%%%%%%%%%%%%%%%%%%%%%%%%
%%%%%%%%%%%%%%%%%%%%%%%%%%%%%%%%%%%%%%%%%%%%%%%%%%%%%%%%%%%%%%%%
\begin{figure}[tb]
\subfigure[]%[$b=0.5$]
{
\label{fig8a}
\begin{minipage}[c]{0.5\textwidth}
\centering\includegraphics[width=9.0cm,height=6.3cm]{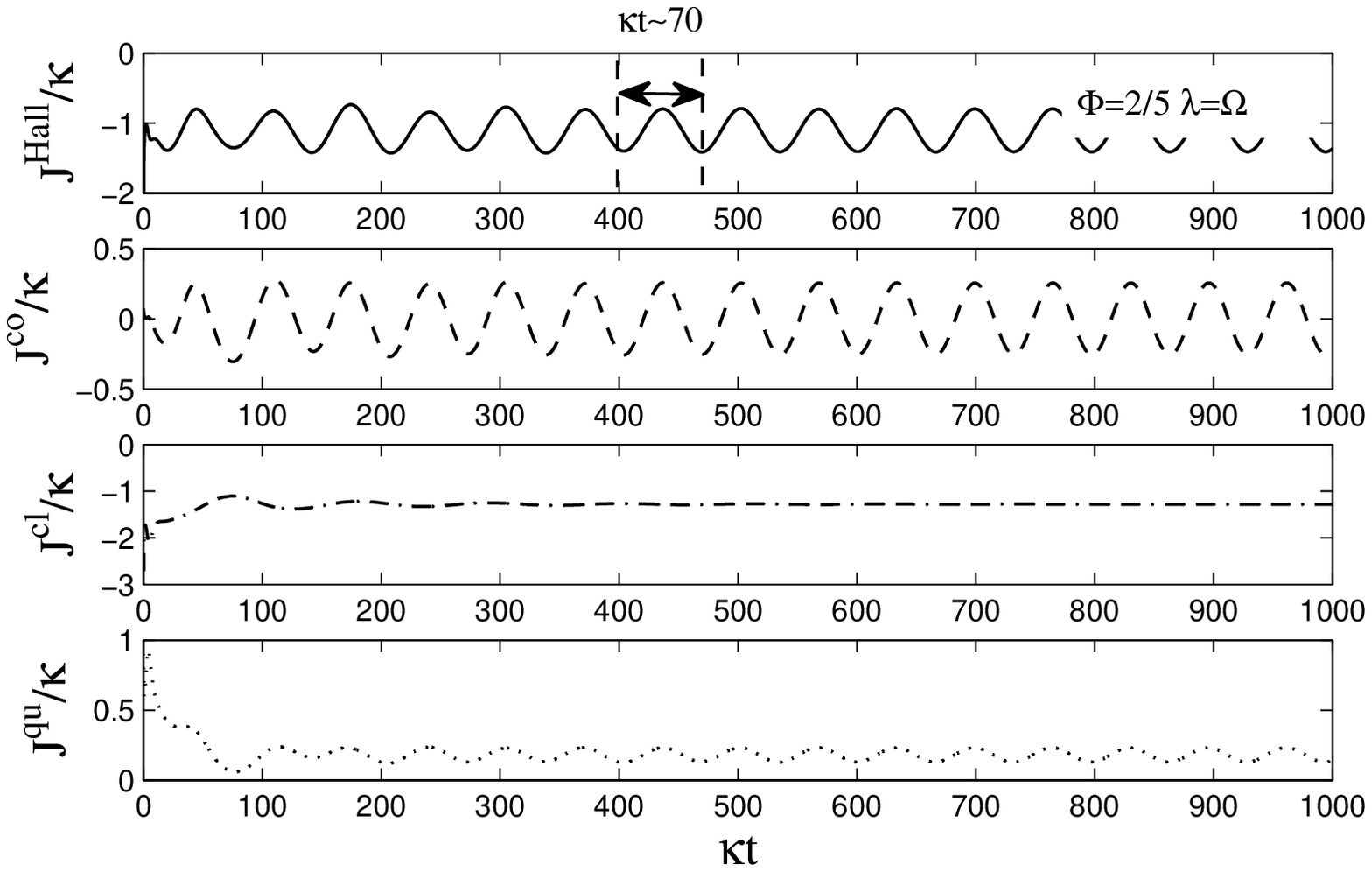}
\end{minipage}}%
\subfigure[]%[$b=3.5$]
{
\label{fig8b}
\begin{minipage}[c]{0.5\textwidth}
\includegraphics[width=9.0cm,height=6.3cm]{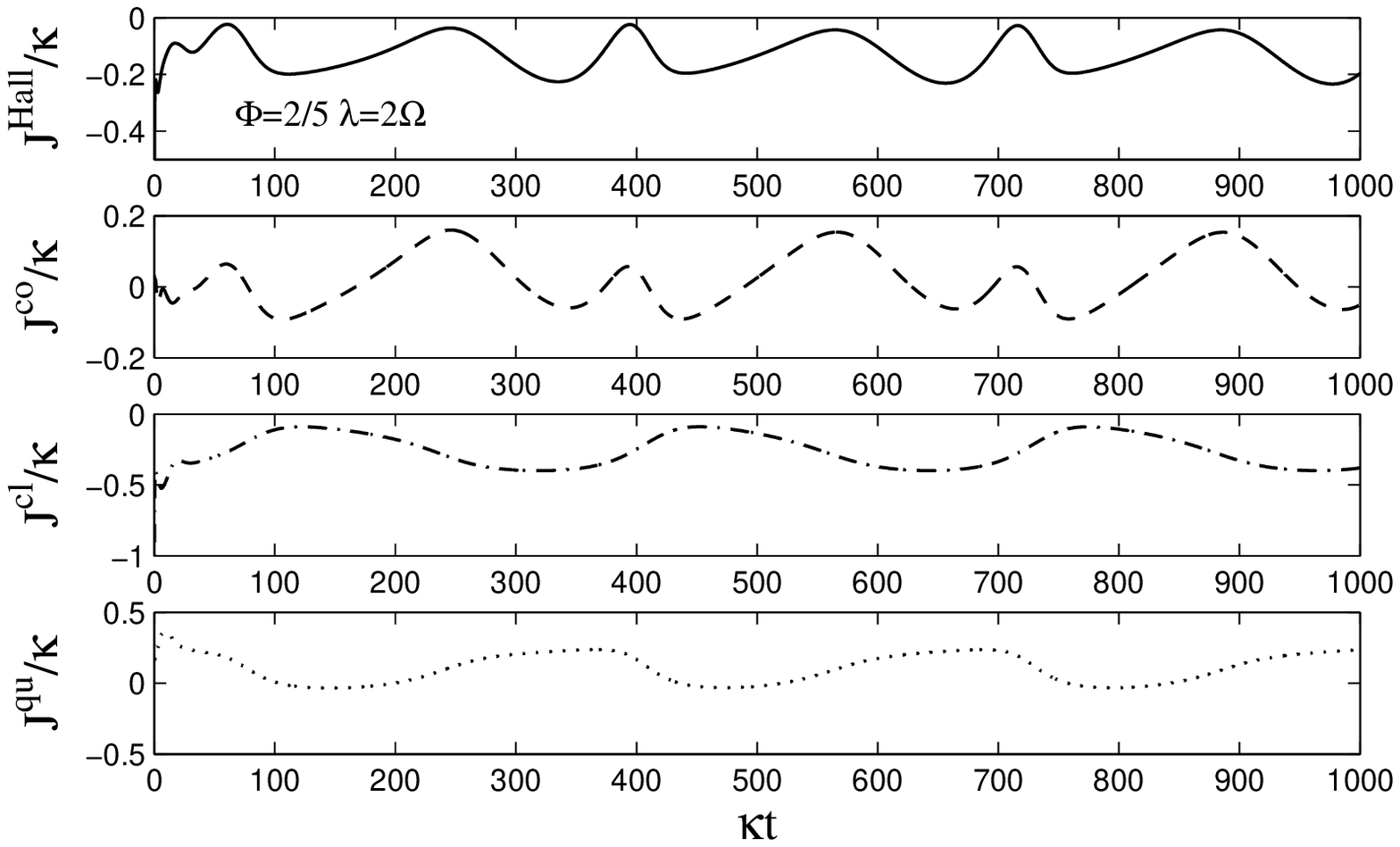}
\end{minipage}}%
\caption{Same as Fig. \ref{fig7}, but for the case of $\Phi=2/5$, in the mean time $\lambda=\Omega$ in Fig. \ref{fig8a} and $\lambda=2\Omega$ in Fig. \ref{fig8b}.}
\label{fig8}
\end{figure}
%%%%%%%%%%%%%%%%%%%%%%%%%%%%%%%%%%%%%%%%%%%%%%%%%%%%%%%
In Figs. \ref{fig6}, \ref{fig7}, and \ref{fig8}, we plot the population of the cavity field and the longitudinal currents as a function of dimensionless time $\kappa t$. Once again we only plot the most representative results. We consider a fixed filling $n=1/4$, and the rational magnetic flux in three cases: $\Phi=0$, $\Phi=1/2$, and $\Phi=2/5$. In the large dissipation regime, other parameters are given by $\lambda/\kappa=0.5$, $\Omega/\kappa=0.5$, and $\Delta/\kappa=0.5$.

For the vanishing magnetic field $\Phi=0$, the cavity population first increases to the maximum and then decreases to zero at $\kappa t=90$. This means that before subjected to the dissipation, a superradiant pulse spontaneously arises. We recognize a pattern that the dynamics can be attributed into the coherent dominant regime characterized by superradiance and the dissipative dominant regime without superradiance. The crossover between these two regimes is obviously a dynamical phase transition. In Fig. \ref{fig7a}, the longitudinal (Hall) current is also time-dependent. It decreases with time until the negative minimum is reached, and then it increases to zero meaning that after a sufficiently long time the center-of-mass motion of atoms will stop. The underlying physics can be easily understood from the rest three panels. Although the semiclassical and quantum transports are both exist in the dissipative dominant regime, where the cavity photons are stemmed from the thermal and quantum fluctuations, they are canceling each other. In the mean time, the coherent current will vanish with the collapse of the superradiance. Therefore there is no persistent current in the steady state. In fact, the arising and vanishing of the superradiance and the atomic current are the manifestation of the dynamical feedback between the cavity field and the center-of-mass motion of atoms. Since the initial condition promotes the atomic transport, the cavity field can be populated which in turn has a positive feedback on the transportation. Accompanied by the cavity loss, the cavity population is declined. When the cavity is empty, the Raman process only supports an unidirectional jump. The superradiant phase depends on the presence of the atomic density wave \cite{Chen}. When the atomic center-of-mass degrees of freedom are freezed by the Pauli exclusion principle, the density wave and the superradiance are all disappeared simultaneously.

The dynamics dramatically changes when the flux is increased to $\Phi=1/2$. The superradiant phase is extended to the dissipative dominant regime, where the curve is flat indicating a constant cavity population. We see in Fig. \ref{fig7b} that the semiclassical and quantum currents have neither same sign nor same amplitude. They can not cancel each other in the dissipative dominant regime. The resulting dissipative atomic hopping forms a density wave with finite amplitude. The feedback of this amplitude on the cavity field gives rise to the persistent cavity population which in turn generates the coherent hopping as shown in the second panel in Fig. \ref{fig7b}. Since the coherent transport is stronger than the other two transports, the total current is dominant by the coherent contribution.

To further illustrate the dynamics in the circumstance which may be more interesting, we demonstrate the case of $\Phi=2/5$ in the lower panel of Fig. \ref{fig6}. The corresponding atomic currents are given in Fig. \ref{fig8}. It is shown that after the initial decrease the total current is negative and periodic with period $\kappa t\sim70$. The former character is cased by the semiclassical transport whose value is negative and reaches the steady value of $J^{cl}\sim-\kappa$. The later character is due to the coherent transport which is periodically changing its plus minus sign. Since the quantum contribution is smaller than the semiclassical one, the dynamics is finally subjected to the semiclassical transport. Different from the previous two cases, the coherent cavity field and the total atomic current are both time-dependent over a very long period of time $\kappa t\approx1000$. Thus one may surmise that the equilibrium can not be reached and the system is always characterized by time-dependent non-equilibrium dynamics. In the above section, the unequal matter-light coupling will alter the fractal and topological properties of the static bulk energy spectrum. To see its influence on the dynamical behavior, we plot the case $\lambda=2\Omega$ in Fig. \ref{fig8b}. Comparing with Fig. \ref{fig8a}, the profile of curves is changed but the behavior is still time-dependent. The semiclassical current that is constant in Fig. \ref{fig8a} becomes periodic. This indicates that the specific values of the couplings only have qualitative influence on the semiclassical transport while quantitative effect in the total atomic current.

From Figs. \ref{fig7} and \ref{fig8}, another dynamical phase transition in the dissipative dominant regime is  surfaced. For the vanishing magnetic flux, the steady state of the cavity field is characterized by thermal and quantum fluctuations. The total atomic transport trends to zero indicating a static atomic spatial distribution. With the increase of the magnetic flux, a dynamical phase transition takes place. When the flux is equal to $2/5$, the coherent cavity field amplitude and the atomic transport are both non-zero and time-dependent. Once the flux exceeds the previous case, the steady state is characterized by the time-independent cavity population with the persistent atomic transport. Therefore, a dynamical phase transition is driven by the magnetic flux. If there is no such flux the artificial electromagnetism is present only in the coherent dominant regime. However, it can be existed in the dissipative dominant regime by a finite flux.
%%%%%%%%%%%%%%%%%%%%%%%%%%%%%%%%%%%%%%%%%%%%%%%%%%%%%%%%%%%%%%%%%%%%%%%%%%%%%%%%%%%%%%%%%%%%%%%%%%%%
\section{conclusion}
\label{con}

In summary, we have investigated the static behavior and the non-equilibrium dynamics of a coupled atom-cavity system. The coupling mechanism is constructed via two Raman processes employing the quantized cavity mode with both traveling- and standing-wave pump beams. This configuration realizes a self-organizing state of the artificial electric and magnetic fields. Static behavior is analyzed by the Harper equation, while the non-equilibrium dynamics is captured by the Keldysh formulism. Analytical and numerical calculations reveal certain new physics:
(I) The static version of our model can be regarded as a generalized Hofstadter model realizing the quantum Hall phase. Its bulk energy spectrum resembles the fractal structure of the Hofstadter butterfly. The edge mode spectrum and the Hall conductivity can be manipulated by setting the values of the matter-light couplings.
(II) With a strong cavity decay, a superradiant pulse with infinitesimal pumping threshold arises. According to the appearance of the superradiance, the dynamics can be divided into two regimes, i.e. the coherent dominant regime with superradiance and the dissipative dominant regime without superradiance. These two regimes are crossed by a dynamical phase transition.
(III) The non-zero longitudinal particle flow indicates the emergence of the artificial electromagnetic field, which results from the coupled dynamics of the cavity photon and the center-of-mass motion of atoms. The feedback between the cavity field and the atomic current shapes the dynamics. Although the Hamiltonian is time-independent in a rotating frame, the behavior of the system may be always time-dependent. The atom-laser couplings has the quantitative influence, while the temporal behavior is determined by the flux of the artificial magnetic field qualitatively.
(IV) In the steady state, the increasing value of the flux drives another dynamical phase transition. For the vanishing flux, the steady state has no superradiance and the particle transport disappears resulting from the cancelation between the thermal and quantum fluctuations. When the flux is increased to $2/5$, the steady state cannot be reached and the system is always characterized by the non-equilibrium dynamics. With the further increase of flux to $1/2$, the superradiance and the atomic flow are both constant in the steady state. In this case, a dynamical equilibrium between fluctuation and dissipation is achieved.

Although the transport measurements are not straightforward to perform in atomic systems as in condensed-matter setups, the Hall conductivity can be detected based on the Streda formula \cite{Kohmoto,Streda}. Moreover, the initial condition of atoms can be launched by a potential wall in a small region located at the lower left corner of the lattice. After releasing the wall, the particle flow is directly observable in in-situ images of the spatial density $\rho_{iji'j'}(t)$ \cite{Goldman1}. In the mean time, counting the statistics of the cavity output field enables the measuring of the superradiant phase in a direct and non-destructive way \cite{Orszag}.
%%%%%%%%%%%%%%%%%%%%%%%%%%%%%%%%%%%%%%%%%%%%%%%%%%%%%%%%%%%%%%%%%%%%%%%%%%%%%%%%%%%%%%%%%%%%%%%%%%%%%%
\section*{ACKNOWLEDGEMENTS}

This work is supported by the National Natural Science Foundation of China (No. 11604240 and No. ), and the Innovation Foundation for Young Teachers of Tianjin University of Science and Technology (No. 2014CXLG22).
%%%%%%%%%%%%%%%%%%%%%%%%%%%%%%%%%%%%%%%%%%%%%%%%%%%%%%%%%%%%%%%%%%%%%%%%%%%%%%%%%%%%%%%%%%%%%%%%%%%%%%%%%
\section*{Appendix: equation of motion}

%%%%%%%%%%%%%%%%%%%%%%%%%%%%%%%%%%%%%%%%%%%%%%
\begin{figure}[tb]
\centering
\includegraphics[width=12cm]{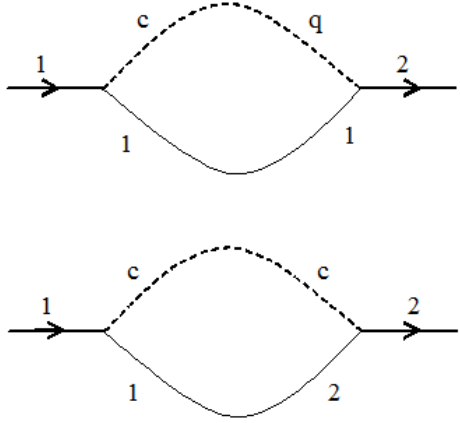}
\caption{Diagrammatic representation of the retarded self-energy $\Sigma^{R}_{i'j'ij}(t_{2},t_{1})$.}
\label{figa1}
\end{figure}
%%%%%%%%%%%%%%%%%%%%%%%%%%%%%%%%%%%%%%%%%%%%%%%%%%%%%%%%%%%
%%%%%%%%%%%%%%%%%%%%%%%%%%%%%%%%%%%%%%%%%%%%%%
\begin{figure}[tb]
\centering
\includegraphics[width=12cm]{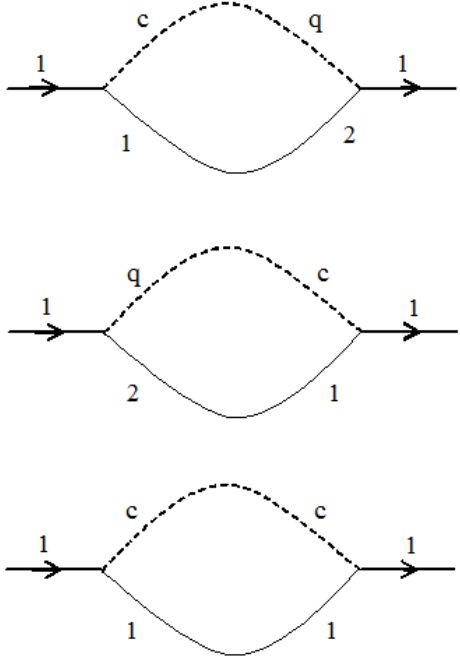}
\caption{Diagrammatic representation of the Keldysh self-energy $\Sigma^{K}_{i'j'ij}(t_{2},t_{1})$.}
\label{figa2}
\end{figure}
%%%%%%%%%%%%%%%%%%%%%%%%%%%%%%%%%%%%%%%%%%%%%%%%%%%%%%%%%%%

The details on the derivation of Eq. (\ref{eomfull}) is the subject of this appendix. The dynamics of the single fermion density matrix, $\rho_{iji'j'}(t)=\langle c^{\dag}_{i,j}(t)c_{i',j'}(t)\rangle$, is related to the Keldysh Green function $G^{K}_{i'j'ij}(t,\omega)$ via
\begin{equation}\label{keldysh1}
\rho_{iji'j'}(t)=\frac{1}{2}\left[\delta_{ii'}\delta_{jj'}-\mathrm{i}\int\frac{d\omega}{2\pi}
G^{K}_{i'j'ij}(t,\omega)\right].
\end{equation}
The kinetics of the Keldysh Green function is governed by \cite{Altland,Rammer}
\begin{equation}\label{eomgk}
[G_{0}^{-1}-\mathrm{Re}\Sigma^{R},G^{K}]_{\circ}+[\mathrm{Re}G^{R},\Sigma^{K}]_{\circ}=
\frac{1}{2}\mathrm{i}\{A,\Sigma^{K}\}_{\circ}-\frac{1}{2}\mathrm{i}\{\Gamma,G^{K}\}_{\circ}.
\end{equation}
Here, $G_{0}$ and $G_{R(A)}$ are the free, retarded (advanced) Green functions, which are defined as
\begin{equation}\label{g0}
G_{0}^{-1}=\mathrm{i}\partial_{t}-H,
\end{equation}
\begin{equation}\label{gr}
G^{R}_{i'j'ij}(t_{2},t_{1})=-\mathrm{i}\theta(t_{2}-t_{1})
\langle\{c_{i',j'}(t_{2}),c^{\dag}_{i,j}(t_{1})\}\rangle=
-\mathrm{i}
\langle\psi^{(1)}_{i',j'}(t_{2})\bar{\psi}^{(2)}_{i,j}(t_{1})\rangle,
\end{equation}
and
\begin{equation}\label{ga}
G^{A}_{i'j'ij}(t_{2},t_{1})=-\mathrm{i}\theta(t_{1}-t_{2})
\langle\{c_{i',j'}(t_{1}),c^{\dag}_{i,j}(t_{2})\}\rangle=
-\mathrm{i}
\langle\psi^{(2)}_{i',j'}(t_{2})\bar{\psi}^{(1)}_{i,j}(t_{1})\rangle.
\end{equation}
In Eqs. (\ref{gr}) and (\ref{ga}), $\{a,b\}=ab+ba$ is the anti-commutator, $\theta(t)$ is the step function, and in the last relations field operators are written on the Keldysh contour. In Eq. (\ref{eomgk}), $\Sigma^{K}$ and $\Sigma^{R(A)}$ are the Keldysh and retarded (advanced) self-energy. The spectrum and lifetime functions are given as
\begin{equation}\label{A}
A_{i'j'ij}(t_{2},t_{1})=\mathrm{i}\left[G^{R}_{i'j'ij}(t_{2},t_{1})-G^{A}_{i'j'ij}(t_{2},t_{1})\right],
\end{equation}
\begin{equation}\label{gamma}
\Gamma_{i'j'ij}(t_{2},t_{1})=\mathrm{i}\left[\Sigma^{R}_{i'j'ij}(t_{2},t_{1})
-\Sigma^{A}_{i'j'ij}(t_{2},t_{1})\right].
\end{equation}
In Eq. (\ref{eomgk}), the commutator (anti-commutator) is defined as $[f_{1},f_{2}]_{\circ}=f_{1}\circ f_{2}-f_{2}\circ f_{1}$ ($\{f_{1},f_{2}\}_{\circ}=f_{1}\circ f_{2}+f_{2}\circ f_{1}$), where $f_{1}\circ f_{2}$ denotes the space-time convolution of the two-point function. Ignoring the self-energy in the left hand side of Eq. (\ref{eomgk}) (i.e., the quasi-particle approximation \cite{Sieberer}), we have that
\begin{equation}\label{eomgk1}
[G_{0}^{-1},G^{K}]_{\circ}=
\frac{1}{2}\mathrm{i}\{A,\Sigma^{K}\}_{\circ}-\frac{1}{2}\mathrm{i}\{\Gamma,G^{K}\}_{\circ}.
\end{equation}
Making the Wigner transformation $f(t,\omega)=\int d\Delta te^{\mathrm{i}\Delta t}f(t,\Delta t)$, where $t=t_{1}+t_{2}$ and $\Delta t=t_{2}-t_{1}$ are the forward and relative time, and under the gradient approximation $(f_{1}\circ f_{2})(t,\omega)=f_{1}(t,\omega)f_{2}(t,\omega)$ \cite{Sieberer}, Eq. (\ref{eomgk1}) yields that
\begin{equation}\label{eomgk2}
\mathrm{i}\partial_{t}G_{i'j'ij}^{K}=-\sum_{ml}\left(h_{i'j'ml}G_{mlij}^{K}-G_{i'j'ml}^{K}h_{mlij}\right).
\end{equation}
In writing Eq. (\ref{eomgk2}), we also ignore the self-energy in the right hand side of Eq. (\ref{eomgk1}) (i.e., the mean-field approximation).
The matrix $h_{iji'j'}$ can be read from the identity $H^{f}=\sum_{iji'j'}c^{\dag}_{i,j}h_{iji'j'}c_{i',j'}$, where $H^{f}$ is the fermionic part of Eq. (\ref{ham1}). We find that
\begin{equation}\label{ham2}
\begin{aligned}
h_{iji'j'}=&-\lambda(\alpha^{\ast}\delta_{i,i'+1}+\alpha\delta_{i+1,i'})\delta_{j,j'}\\
&-\Omega(\alpha^{\ast}e^{\mathrm{i}\phi i}\delta_{j,j'+1}
+\alpha e^{-\mathrm{i}\phi i}\delta_{j+1,j'})\delta_{i,i'}.
\end{aligned}
\end{equation}
Inserting the above equation into Eq. (\ref{eomgk2}), performing the sum over indexes $m$ and $l$, and using the identity $\int\frac{d\omega}{2\pi}
G^{K}_{i'j'ij}(t,\omega)=\mathrm{i}\left[2\rho_{iji'j'}(t)-\delta_{ii'}\delta_{jj'}\right]$, we obtain the equation of motion of $\rho_{iji'j'}(t)$ as
\begin{equation}\label{eom1}
\begin{aligned}
\partial_{t}\rho_{iji'j'}(t)=&\mathrm{i}\lambda(\alpha^{\ast}\rho_{iji'-1j'}+\alpha\rho_{iji'+1j'}
-\alpha^{\ast}\rho_{i+1ji'j'}-\alpha\rho_{i-1ji'j'})\\
&+\mathrm{i}\Omega(\alpha^{\ast}e^{\mathrm{i}\phi i'}\rho_{iji'j'-1}
+\alpha e^{-\mathrm{i}\phi i'}\rho_{iji'j'+1}
-\alpha^{\ast}e^{\mathrm{i}\phi i}\rho_{ij+1i'j'}
-\alpha e^{-\mathrm{i}\phi i}\rho_{ij-1i'j'}).
\end{aligned}
\end{equation}
The above equation together with Eq. (\ref{eomp}) determines the coherent dynamics of the system.

To incorporate the fluctuations of the cavity field into the dynamics, one should re-derive the equation of motion started from Eq. (\ref{eomgk1}). For our system, its fluctuation part reads
\begin{equation}\label{eomgk3}
\begin{aligned}
\mathrm{i}\partial_{t}G_{i'j'ij}^{K}(t,\omega)=&\frac{\mathrm{i}}{2}\sum_{ml}\left[
A_{i'j'ml}(t,\omega)\Sigma^{K}_{mlij}(t,\omega)+\Sigma^{K}_{i'j'ml}(t,\omega)A_{mlij}(t,\omega)\right]\\
&-\frac{\mathrm{i}}{2}\sum_{ml}\left[
\Gamma_{i'j'ml}(t,\omega)G^{K}_{mlij}(t,\omega)+G^{K}_{i'j'ml}(t,\omega)\Gamma_{mlij}(t,\omega)\right].
\end{aligned}
\end{equation}
To calculate the self-energy in the diagrammatic method, the vertex function should be extracted from the Keldysh action that is associated with the interacting part of the Hamiltonian (\ref{ham1}). After some algebra, it turns out that
\begin{equation}\label{kaction}
S_{I}=\frac{1}{\sqrt{2}}\sum_{iji'j'}\bar{\Psi}_{ij}\left[(\phi_{c}^{\ast}v_{1}(iji'j')
+\phi_{c}v_{2}(iji'j'))\sigma_{0}
+(\phi_{q}^{\ast}v_{1}(iji'j')+\phi_{q}v_{2}(iji'j'))\sigma_{1}\right]\Psi_{i'j'}.
\end{equation}
Here, $\phi_{c}$, $\phi_{q}$, $\bar{\Psi}_{ij}=\left(\bar{\psi}^{(1)}_{ij},\bar{\psi}^{(2)}_{ij}\right)$, and $\Psi_{i'j'}=\left(\psi^{(1)}_{i'j'},\psi^{(2)}_{i'j'}\right)^{\mathrm{T}}$ denote the boson and fermion fields on the Keldysh contour respectively. The vertex functions are given as $v_{1}(iji'j')=\lambda\delta_{ii'+1}\delta_{jj'}+\Omega e^{\mathrm{i}\phi i}\delta_{ii'}\delta_{jj'+1}$ and $v_{2}(iji'j')=\lambda\delta_{i+1i'}\delta_{jj'}+\Omega e^{-\mathrm{i}\phi i}\delta_{ii'}\delta_{j+1j'}$.
As depicted in Fig. \ref{figa1}, the retarded self-energy derived by the common diagrammatic rule reads
\begin{equation}\label{SigmaR}
\begin{aligned}
\Sigma^{R}_{i'j'ij}(t_{2},t_{1})=&\frac{\mathrm{i}}{2}\sum_{mlm'l'}\left[
v_{11}v_{22}D^{R}(t_{2},t_{1})G^{K}_{mlm'l'}(t_{2},t_{1})
+v_{21}v_{12}D^{A}(t_{1},t_{2})G^{K}_{mlm'l'}(t_{2},t_{1})\right]\\
&+\frac{\mathrm{i}}{2}\sum_{mlm'l'}\left[
v_{11}v_{22}D^{K}(t_{2},t_{1})G^{R}_{mlm'l'}(t_{2},t_{1})
+v_{21}v_{12}D^{K}(t_{1},t_{2})G^{R}_{mlm'l'}(t_{2},t_{1})\right],
\end{aligned}
\end{equation}
where $v_{11}=v_{1}(i'j'ml)$, $v_{22}=v_{2}(m'l'ij)$, $v_{21}=v_{2}(i'j'ml)$, and $v_{12}=v_{1}(m'l'ij)$ are the abbreviations for the vertex functions with different arguments. Here, $D^{K}$, $D^{R(A)}$ are the Keldysh, retarded (advanced) Green function for photons. The advanced self-energy is the hermitian conjugate of Eq. (\ref{SigmaR}) which reads
\begin{equation}\label{SigmaA}
\begin{aligned}
\Sigma^{A}_{i'j'ij}(t_{2},t_{1})=&\frac{\mathrm{i}}{2}\sum_{mlm'l'}\left[
v_{11}v_{22}D^{A}(t_{2},t_{1})G^{K}_{mlm'l'}(t_{2},t_{1})
+v_{21}v_{12}D^{R}(t_{1},t_{2})G^{K}_{mlm'l'}(t_{2},t_{1})\right]\\
&+\frac{\mathrm{i}}{2}\sum_{mlm'l'}\left[
v_{11}v_{22}D^{K}(t_{2},t_{1})G^{A}_{mlm'l'}(t_{2},t_{1})
+v_{21}v_{12}D^{K}(t_{1},t_{2})G^{A}_{mlm'l'}(t_{2},t_{1})\right].
\end{aligned}
\end{equation}
According to Fig. \ref{figa2}, the Keldysh self-energy reads
\begin{equation}\label{SigmaK}
\begin{aligned}
\Sigma^{K}_{i'j'ij}(t_{2},t_{1})=&\frac{\mathrm{i}}{2}\sum_{mlm'l'}\left[
v_{11}v_{22}D^{K}(t_{2},t_{1})G^{K}_{mlm'l'}(t_{2},t_{1})
+v_{21}v_{12}D^{K}(t_{1},t_{2})G^{K}_{mlm'l'}(t_{2},t_{1})\right]\\
&+\frac{\mathrm{i}}{2}\sum_{mlm'l'}\left[
v_{11}v_{22}D^{A}(t_{2},t_{1})G^{A}_{mlm'l'}(t_{2},t_{1})
+v_{21}v_{12}D^{R}(t_{1},t_{2})G^{A}_{mlm'l'}(t_{2},t_{1})\right]\\
&+\frac{\mathrm{i}}{2}\sum_{mlm'l'}\left[
v_{11}v_{22}D^{R}(t_{2},t_{1})G^{R}_{mlm'l'}(t_{2},t_{1})
+v_{21}v_{12}D^{A}(t_{1},t_{2})G^{R}_{mlm'l'}(t_{2},t_{1})\right].
\end{aligned}
\end{equation}
To proceed, we transform the life-time function to the Wigner representation
\begin{equation}\label{Gamma}
\begin{aligned}
&\Gamma_{i'j'ij}(t,\omega)=\int d\Delta te^{\mathrm{i}\omega t}\Gamma_{i'j'ij}(t,\Delta t)\\
&=\mathrm{i}\int d\Delta te^{\mathrm{i}\omega t}\frac{\mathrm{i}}{2}\sum_{mlm'l'}\left[
\tilde{v}_{1}(D^{R}-D^{A})(t_{2},t_{1})G^{K}_{mlm'l'}(t_{2},t_{1})
+\tilde{v}_{2}(D^{A}-D^{R})(t_{1},t_{2})G^{K}_{mlm'l'}(t_{2},t_{1})\right]\\
&+\frac{\mathrm{i}}{2}\sum_{fi}\left[
\tilde{v}_{1}D^{K}(t_{2},t_{1})(G^{R}-G^{A})_{mlm'l'}(t_{2},t_{1})
+\tilde{v}_{2}D^{K}(t_{1},t_{2})(G^{R}-G^{A})_{mlm'l'}(t_{2},t_{1})\right],
\end{aligned}
\end{equation}
where $\tilde{v}_{1}=v_{11}v_{22}=v_{1}(i'j'ml)v_{2}(m'l'ij)$ and $\tilde{v}_{2}=v_{21}v_{12}=v_{2}(i'j'ml)v_{1}(m'l'ij)$ are the abbreviations. Inserting the Wigner transform of the Green functions
$G(t,\Delta t)=\int\frac{d\omega'}{2\pi}e^{-\mathrm{i}\omega'\Delta t}G(t,\omega')$ into Eq. (\ref{Gamma}) and integrating over the relative time, we obtain that
\begin{equation}\label{Gamma1}
\begin{aligned}
&\Gamma_{i'j'ij}(t,\omega)=
-\frac{1}{2}\int\frac{d\nu}{2\pi}\sum_{mlm'l'}\left[
\tilde{v}_{1}(D^{R}-D^{A})(t,\nu)G^{K}_{mlm'l'}(t,\omega-\nu)
+\tilde{v}_{2}(D^{A}-D^{R})(t,\nu)G^{K}_{mlm'l'}(t,\omega+\nu)\right]\\
&+\sum_{mlm'l'}\left[
\tilde{v}_{1}D^{K}(t,\nu)(G^{R}-G^{A})_{mlm'l'}(t,\omega-\nu)
+\tilde{v}_{2}D^{K}(t,\nu)(G^{R}-G^{A})_{mlm'l'}(t,\omega+\nu)\right].
\end{aligned}
\end{equation}
In the large dissipation limit, we neglect the photon self-energy, and approximate
\begin{equation}\label{approx1}
\begin{aligned}
D^{K}(t,\nu)&=|D^{R}(t,\nu)|^{2}\left[\Pi^{K}(t,\nu)-2\mathrm{i}\kappa\right]\\
&\approx\frac{-2\mathrm{i}\kappa}{\Delta^{2}+\kappa^{2}},
\end{aligned}
\end{equation}
and
\begin{equation}\label{approx2}
\begin{aligned}
(D^{R}-D^{A})(t,\nu)&=|D^{R}(t,\nu)|^{2}\left[\Pi^{R}(t,\nu)-\Pi^{A}(t,\nu)-2\mathrm{i}\kappa\right]\\
&\approx\frac{-2\mathrm{i}\kappa}{\Delta^{2}+\kappa^{2}}.
\end{aligned}
\end{equation}
Inserting the above two equations into Eq. (\ref{Gamma1}) and using the identity $G^{R}-G^{A}=-\mathrm{i}A$, we have that
\begin{equation}\label{Gamma2}
\begin{aligned}
\Gamma_{i'j'ij}(t,\omega)=&
\frac{\mathrm{i}\kappa}{\Delta^{2}+\kappa^{2}}\int\frac{d\nu}{2\pi}\{\sum_{mlm'l'}\left[
\tilde{v}_{1}G^{K}_{mlm'l'}(t,\omega-\nu)
-\tilde{v}_{2}G^{K}_{mlm'l'}(t,\omega+\nu)\right]\\
&-\mathrm{i}\sum_{mlm'l'}\left[
\tilde{v}_{1}A_{mlm'l'}(t,\omega-\nu)
+\tilde{v}_{2}A_{mlm'l'}(t,\omega+\nu)\right]\}.
\end{aligned}
\end{equation}
The Wigner transform of the Keldysh self-energy (\ref{SigmaK}) can be work out by the similar procedure. After a straightforward calculation, it yields that
\begin{equation}\label{SigmaK1}
\begin{aligned}
\Sigma^{K}_{i'j'ij}(t,\omega)=&\frac{\kappa}{\Delta^{2}+\kappa^{2}}\int\frac{d\nu}{2\pi}\{
\sum_{mlm'l'}\mathrm{i}\left[
\tilde{v}_{2}A_{mlm'l'}(t,\omega+\nu)-\tilde{v}_{1}A_{mlm'l'}(t,\omega-\nu)\right]\\
&+\left[\tilde{v}_{1}G^{K}_{mlm'l'}(t,\omega-\nu)
+\tilde{v}_{2}G^{K}_{mlm'l'}(t,\omega+\nu)\right]\}.
\end{aligned}
\end{equation}
Substituting Eqs. (\ref{Gamma2}) and (\ref{SigmaK1}) into Eq. (\ref{eomgk3}), it can be rewritten as
\begin{equation}\label{eom3}
\begin{aligned}
\partial_{t}\rho_{iji'j'}=&-\frac{\kappa}{\Delta^{2}+\kappa^{2}}\sum_{\alpha\beta ml}\left[
v_{1}(i'j'\alpha\beta)v_{2}(\alpha\beta ml)\rho_{ijml}
+v_{1}(ml\alpha\beta)v_{2}(\alpha\beta ij)\rho_{mli'j'}\right]\\
&+\frac{2\kappa}{\Delta^{2}+\kappa^{2}}\sum_{\alpha\beta\alpha'\beta'}
v_{2}(i'j'\alpha\beta)v_{1}(\alpha'\beta' ij)
\rho_{\alpha'\beta'\alpha\beta}\\
&+\frac{\kappa}{\Delta^{2}+\kappa^{2}}\sum_{\alpha\beta\alpha'\beta'}\sum_{ml}\left[
(\bar{v}_{1}-\bar{v}_{2})\rho_{\alpha'\beta'\alpha\beta}\rho_{ijml}
+(\bar{v}'_{1}-\bar{v}'_{2})\rho_{mli'j'}\rho_{\alpha'\beta'\alpha\beta}\right],
\end{aligned}
\end{equation}
where the abbreviations are given by $\bar{v}_{1}=v_{1}(i'j'\alpha\beta)v_{2}(\alpha'\beta' ml)$, $\bar{v}_{2}=v_{2}(i'j'\alpha\beta)v_{1}(\alpha'\beta' ml)$, $\bar{v}'_{1}=v_{1}(ml\alpha\beta)v_{2}(\alpha'\beta' ij)$, and $\bar{v}'_{2}=v_{2}(ml\alpha\beta)v_{1}(\alpha'\beta' ij)$. Together with the coherent part given in Eq. (\ref{eom1}), we finally obtain Eq. (\ref{eomfull}) in the main text.
%%%%%%%%%%%%%%%%%%%%%%%%%%%%%%%%%%%%%%%%%%%%%%%%%%%%%%%%%%%%%%%%%%%%%%%%%%%%%%%%%%%%%%%%%%%%%%%%%%%%%%%%%

\end{document}